\documentclass[12pt]{article}
\usepackage[latin1]{inputenc}
\usepackage{graphicx}
\newcommand{\Ncal}{{\cal N}}
\newcommand{\drm}{{\rm d}}

\newcommand{\pa}{\partial}

\newcommand{\text}{\rm}

\newcommand{\ug}{ \; = \; }

\newcommand{\bb}{\begin{equation}}
\newcommand{\ee}{\end{equation}}
\newcommand{\bega}{\begin{eqnarray}}
\newcommand{\ega}{\end{eqnarray}}
\newcommand{\begae}{\begin{eqnarray*}}
\newcommand{\egae}{\end{eqnarray*}}

\newcommand{\h}{\hspace*{4ex}}
\newcommand{\dis}{\displaystyle}

\newcommand{\om}{\omega}

\newcommand{\cent}{\centerline}
\newcommand{\vs}{\vspace*}

\begin{document}

\baselineskip 0.8cm

\begin{center}

{\large {\bf Creating Light-Made Waveguides with Structured Optical Beams in Nonlinear Kerr Media.}}



\end{center}

\vs{3mm}

\cent{ M. Zamboni-Rached\footnote{E-mail addresses for contacts:
mzamboni@decom.fee.unicamp.br}}

\vs{0.15 cm}

\centerline{{\em School of Electrical and Computer Engineering, University of Campinas, Campinas, S\~ao Paulo, Brazil}}

\vs{0.2 cm}

\cent{K. Z. N\'obrega}

\vs{0.15 cm}

\centerline{{\em Department of Teleinformatics Engineering, Federal University of Cear\'a (UFC), Fortaleza, Ceara, Brazil}}

\vs{0.2 cm}

\cent{Mo Mojahedi}

\vs{0.15 cm}

\centerline{{\em The Edward S. Rogers Department of Electrical and Computer Engineering, University of Toronto,}}

\centerline{{\em 10 King's College Road, Toronto, Ontario M5S 3G4, Canada}}

\vs{0.4 cm}

{\bf Abstract  \ --} \ While in linear optics the subject of structured light has been a fruitful field of both theoretical and applied research, its development in the arena of nonlinear optics has been underexplored. In this paper, we construct Frozen-Wave-type structured optical beams in Kerr nonlinear media, emphasizing the self-defocusing case, and use them to guide and control Gaussian optical beams. The results presented in this study support the expectation that structured light in nonlinear media can open new venues of theoretical research and applications, particularly in the realms of light controlling light and for all-optical photonics.

\section{Introduction}

\h Research into in structured light has made remarkable progress in recent years \cite{livro1,livro2}. For example, it has been shown that it is possible to make beam resistant to concurrent effects of diffraction and attenuation \cite{fw1,fw1b,fw1c,fw1d}, to generate beams with exotic rectilinear or curved paths \cite{feixe_curvo,yousuf1,yousuf2}, to exercise control over beam's orbital angular momentum density, polarization, and wavelength \cite{fw2,forbes}, and to use structured light in refractometery \cite{ahmed} and optical tweezers \cite{pinca}. At the core of structured light theory is linearity, since it is through the appropriate superposition of free space propagation modes (such as Bessel beams and/or plane waves) that the modeling of light beams is made possible.

\h Given its dependence on the superposition principle, the topic of structured light has been only scarcely explored in the realm of nonlinear optics \cite{livro3}, which is perhaps one of the most important and fascinating fields of research in physics and engineering, with fundamental applications in optical communications, optical switches, sensors, optical computing, etc. Structured optical fields in the context of nonlinear optics could open new venues in theory and applications, as it is reported in \cite{forbes}.
In this work, we address two important points: we demonstrate the feasibility of obtaining Frozen Wave (FW) type structured optical beams in bulk Kerr media for both positive and negative nonlinear coefficient $n_2$ and, at same time, we explore the potential use of these structured light beams as waveguides, combiners, and splitters for Gaussian beams.

\section{A concise explanation of Frozen Waves in linear media and an examination of the nonlinear scenario.}

\h The FW method \cite{fw1,fw3} allows us to model an optical beam in a homogeneous and linear medium with refractive index $n=n_l$ in such a way that its intensity is concentrated and longitudinally modulated on demand over a surface, whose cross-section radius can also be controlled along the propagation direction, which here is chosen to be the positive $z$-direction.

\h In mathematical terms, we require that, within the interval $0\leq z\leq L$, $|\Psi(\rho=r(z),\phi,z,t)_{LFW}|^2 = [f(z)]^2$, where $r(z)$ represents the desired variable radius (along the z-axis) of the cross section of the surface on which the optical field is concentrated, with its longitudinal intensity pattern selectively determined through the square of the function $f(z)$. The subscript "LFW" denotes "Linear Frozen Wave," emphasizing that the solution in question is applicable to a linear medium.

\h The FW method enables us to generate a beam with the aforementioned properties by expressing it as a superposition of co-propagating Bessel beams of order $\nu$ along the positive $z$-direction \cite{fw3}:

\begin{equation} \label{FW1}
        \Psi(\rho,\phi,z,t)_{LFW} \ug \Ncal_{\nu}\,e^{-i\omega t}\sum^{N}_{m=-N}A_{m}
        J_{\nu}(k_{\rho m}\rho)e^{i\nu\phi}e^{ik_{z m} z}\,\,,
        \end{equation}
    with the longitudinal wave numbers chosen as $k_{z m} \ug Q + \frac{2\pi}{L}m$, while the transverse wave numbers are given by $k_{\rho m} = \sqrt{k^2 - k_{z m}^2}\,$, where $k = n_l 2\pi/\lambda$ is the wave number in the medium and $\lambda$ is the wavelength in free space. The complex coefficients of the superposition are calculated from

\begin{equation} \label{An}
        A_{m} = \frac{1}{L} \int^{L}_{0}F(z) e^{-i\frac{2\pi}{L}m\,z}
        \drm z \;\;,\;\; {\rm with } \;\; F(z) = f(z) e^{i g(z)}\,\, ,
        \end{equation}
        where $F(z)$ is referred to as the morphological function, and the phase function $g(z)$ is related to $r(z)$ according to

 \bb r(z) = \zeta_{\nu}/\sqrt{k^2 - \left(Q + \pa g/\pa z\right)^2}\,\,,
    \label{g}\ee
    where, for $\nu \neq 0$, $r(z)$ represents the desired variable cross-section radius (along the $z$-direction) of the surface over which the hollow beam field $\Psi_{LFW}$ is concentrated, and $\zeta_{\nu}$ denotes the value of $\zeta$ at which $J_{\nu}(\zeta)$ reaches its maximum value. In the case of $\nu=0$, we have $\zeta_0 \approx 2.405$, and $r(z)$ represents the desired variable spot radius of a pencil-like beam along the $z$-direction.

    \h It should be clear that Eq.(\ref{g}) enables us to determine the phase-function $g(z)$ that yields the desired radius $r(z)$. For a constant $r(z)$, setting $g(z)=0$, results in $r = \zeta_{\nu}/\sqrt{k^2-Q^2}$, which depends on the FW order, wavelength, and the parameter $Q$, referred to here as the FW's $Q$ parameter.

\h As examples of application of FWs in linear medium, we show two fourth-order FW beams in a medium with $n_l = 1.3$, both defined in the interval $0 \leq z \leq L = 20$cm. In the first example, we generate a hollow beam that is concentrated on a cylindrical surface with a constant radius $r(z)=26.5\mu$m, which implies $g=0$ and, for $\nu=4$, $Q=0.99992k$. The desired longitudinal intensity pattern follows a triple on-off configuration, which is determined by the function $f(z) = [H(z) - H(z-4\rm{cm})] + [H(z-5\rm{cm}) - H(z-9\rm{cm})] + [H(z-10\rm{cm}) - H(z-15\rm{cm})]$, where $H(\cdot)$ is the Heavised function. The coefficients $A_n$ can be obtained from Eq.(\ref{An}). Figure (1a) shows the resulting beam intensity, calculated using Eq.(\ref{FW1}) with $N=40$.

\h As the second example, we select a taper as the surface on which the optical beam is concentrated, extending from $z=0$ to $z=10$cm and with the cross-sectional radius varying according to $r(z) = 5.32/\sqrt{k^2 - (Q+ab\cos(bz))^2}$, with $a=18.95$ and $b=31.42\rm{m}^{-1}$. In this case, we wish the optical beam to have a constant and unit intensity pattern, i.e., $f(z) = 1$ (arbitrary units). Using Eq.(\ref{g}), we determine that $g(z)=a\sin(bz)$, and the complex amplitudes are obtained through Eq.(\ref{An}). Figure (1b) depicts the resulting beam intensity, calculated using Eq.(\ref{FW1}) with $N=24$.

\begin{figure}[ht!]
\centering\includegraphics[width=15cm]{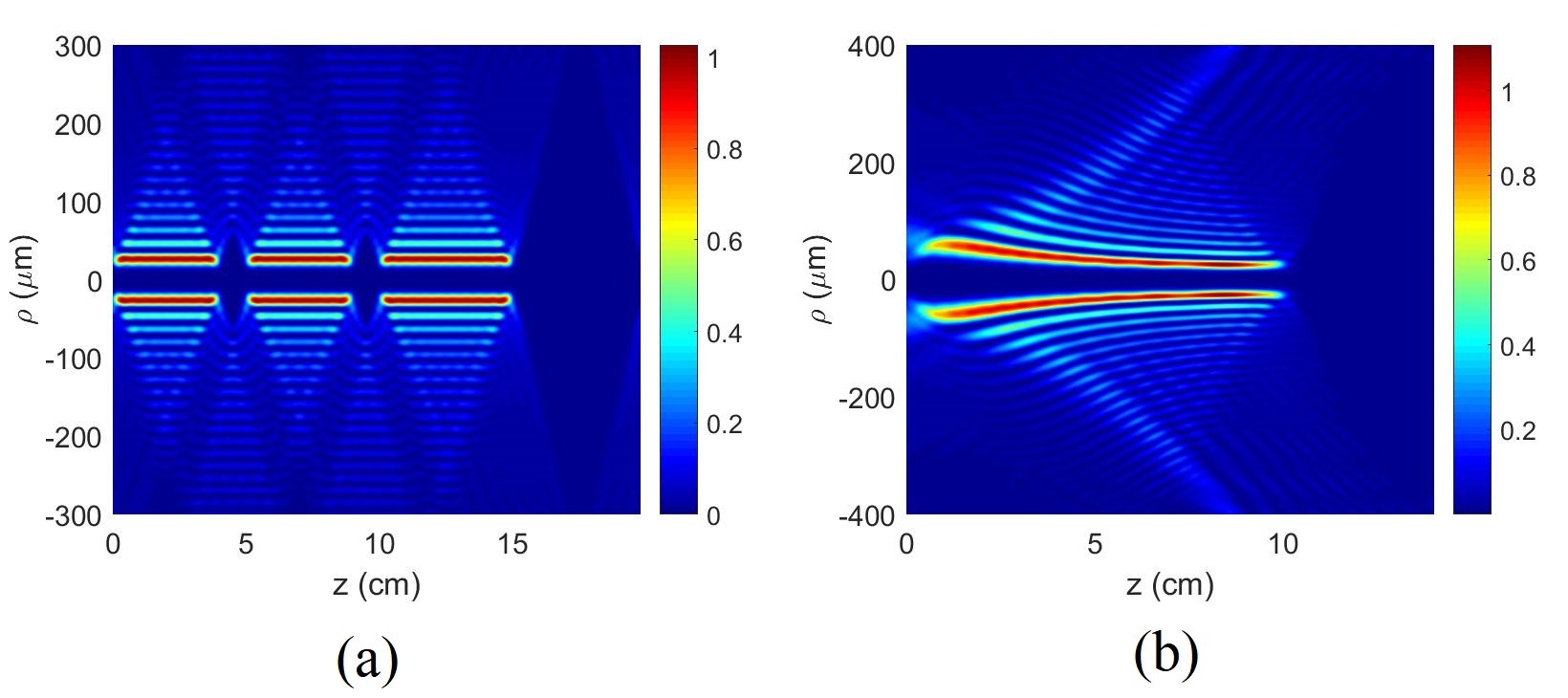}
\caption{Intensities of two fourth-order hollow FWs in a linear medium with a refractive index of $n_l = 1.3$, namely: a) a triple on-off configuration, and b) a taper configuration.}
\end{figure}

\h While the FW method was originally developed for linear media \cite{fw1,fw3}, the unique characteristic of diffraction resistance combined with the possibility of spatially structuring the optical beam present intriguing possibilities when considering a Kerr-type non-linear medium. In the case where the nonlinear coefficient is negative ($n2<0$), it is known that an optical beam experiences self-defocusing, which exacerbates the effect of diffraction, causing the transverse field pattern to spread spatially. Given that FW beams are diffraction resistant, it is natural to speculate that they would also demonstrate greater resistance to self-defocusing as compared to ordinary beams.

\h Conversely, when the nonlinear coefficient is positive ($n2>0$), self-focusing occurs. While this effect can counterbalance the diffraction for 1D beams, resulting in spatial solitons, it leads to the collapse of 2D beams. In this scenario, the use of FWs remains intriguing if we consider the possibility of engineering the longitudinal intensity pattern of the 2D beam in such a way to avoid its collapse within the region of interest.

\h Throughout this paper, the aforementioned possibilities will be confirmed within the context of our main objective, which is to utilize FWs in a Kerr nonlinear medium to create waveguides made of light for guiding Gaussian beams.

\section{Frozen Waves acting as waveguides in nonlinear Kerr media.}

\h The refractive index in a nonlinear Kerr medium depends on the light beam's intensity, i.e., $n = n_l + n_2I$, where $n_l$ represents the linear refractive index, $n_2$ is the nonlinear coefficient, $I = n_l |\Psi|^2/2\eta_0 [W/m^2]$ is the intensity, $\Psi$ denotes the electric field amplitude of the beam, and $\eta_0$ is the vacuum impedance. For materials with a negative $n_2$, regions with higher field strengths exhibit a lower refractive index, while the opposite effect occurs for materials with a positive $n_2$. Therefore, it becomes conceivable that in a medium with $n_2<0$ ($n_2>0$), a hollow optical beam (a pencil-like beam) with a suitable intensity and capable of preserving its transverse pattern over a certain distance, can induce a refractive index distribution that enables the guidance of a Gaussian beam for distances significantly greater than its diffraction length (Rayleigh distance) $L_{diff}=\sqrt{3}/2 k \Delta\rho^2$, where $k= n_l 2 \pi/\lambda$ and $\Delta\rho$ is the beam waist radius. In fact, in \cite{Bessel_NL} the authors investigated the utilization of a simple higher-order Bessel beam in a Kerr medium with $n_2<0$ to successfully guide a Gaussian beam within it.

\h In this section, we will explore these possibilities further by employing FW beams in a Kerr medium to construct guiding structures with intricate patterns far beyond a basic uniform guide. More precisely, exact solutions of FWs shaped according to desired geometries in the linear medium (with $n = n_l$) will be evaluated at $z=0$ and the result will provide the initial excitation (initial field) for the corresponding FWs in the Kerr medium under consideration. Following this procedure, we will be able to obtain in a Kerr medium with $n_2<0$ guiding structures made of light in triple on-off, taper, combiner, and beam splitter formats. To demonstrate the guiding capabilities of such optical structures, Gaussian beams will be inject into these waveguides made of light and we will see that they are actually guided over significantly longer distances compared to their diffraction lengths.

\h When $n_2>0$, the creation of these light-based guiding structures presents greater challenges due to the critical self-focusing effects that a FW experiences, which can cause the beam's collapse. Nevertheless, by carefully manipulating the FW's longitudinal intensity pattern, it is possible to prevent such collapse within the desired region and achieve interesting results. In this case, we will use the FWs to create a light-based guiding structure in the form of an elbow and guide a Gaussian beam along it.

\h In this work, we will adopt a linear refractive index of $n_l = 1.3$. To avoid the need for excessively high power levels to induce the required refractive index contrasts for guiding the Gaussian beams, we will use $|n_2| = 10^{-9} \text{m}^2/\text{W}$, a value comparable to that used in Ref. \cite{{valor_n2}}, which was obtained from a colloid of biosynthesized gold nanoparticles.

\h Considering optical fields associated with the structured beams (FWs) and Gaussian beams, with wavelengths $\lambda_1 = 514.5$nm and $\lambda_2 = 632.8$nm, respectively, given as $\mathbf{E_{FW}} = \Psi_1(x,y,z)e^{i(k_1 z - \om_1 t)}\hat{x}$ and $\mathbf{E_{Gauss}} = \Psi_2(x,y,z)e^{i(k_2 z - \om_2 t)}\hat{x}$, with $k_1 = n_l 2\pi/\lambda_1$ and $k_2 = n_l 2\pi/\lambda_2$, the coupled nonlinear Schroedinger equations \cite{livro3} for the envelopes $\Psi_1(x,y,z)$ and $\Psi_2(x,y,z)$ are given by

\bb \left\{\begin{array}{l}
  \dis{\frac{\pa \Psi_1}{\pa z}} \ug \dis{\frac{i}{2k_1}} \nabla^2_{T}\Psi_1 + \dis{\frac{ik_1 n_2}{2\eta_0}}\dis{(|\Psi_1|^2 \Psi_1 + 2|\Psi_2|^2 \Psi_1)} \\
 \\
\dis{\frac{\pa \Psi_2}{\pa z}} \ug \dis{\frac{i}{2k_2}} \nabla^2_{T}\Psi_2 + \dis{\frac{ik_2 n_2}{2\eta_0}}\dis{(|\Psi_2|^2 \Psi_2 + 2|\Psi_1|^2 \Psi_2)} \,\, .
\end{array} \right. \label{NLS}
 \ee
 
 \

\h Numerical solutions for the system given by Eq.(\ref{NLS}) will be obtained by using the split-step Fourier method, starting from the initial field values on the $z=0$ plane, namely $\Psi_1(x,y,z=0)$ and $\Psi_2(x,y,z=0)$. As previously mentioned, the initial value of $\Psi_1$ will be derived from the FW configuration designed with the desired pattern in the linear medium (with refractive index $n = n_l = 1.3$), and evaluated at $z=0$. Thus, we have $\Psi_1(x,y,0) = \Psi_{LFW}(x,y,0)$ or, if an array of FWs is needed, $\Psi_1(x,y,0) = \sum_{j}{\Psi_{LFW}^{(j)}(x,y,0)}$. On the other hand, $\Psi_2(x,y,0)$ will be represented by a single or a superposition of Gaussian functions.

\h For the examples provided in this paper, the initial amplitude of each Gaussian beam (i.e., $A_g$) will be chosen such that the corresponding nonlinear lengths are 10 times greater than the diffraction lengths, i.e., $L_{nl} = 10 L_{diff}$, and this is achieved by setting $A_g = \sqrt{\left|2\eta_0/(L_{nl} n_2k_2)\right|}$. The amplitudes of the FW beams will be chosen in order to produce the necessary contrast in the refractive index so that the Gaussian beams considered are guided.

\subsection{Examples for $n_2<0$.}

{\bf A triple on-off guiding structure made of light}

\h Here, we will employ a structured hollow beam with a radius of $26.5\mu$m, designed as a triple on-off pattern, to guide a Gaussian beam within its hollow regions. For this purpose, we will use the FW solution from the first example of Section 2 with an amplitude that is a multiple of the Gaussian amplitude given by $2\sqrt{3}A_g$ evaluated at $z=0$, as the initial field $\Psi_1$ in the nonlinear medium. A Gaussian beam will be injected into the first light cylinder, in other words, we will consider as the initial field (i.e., at $z=0$) of $\Psi_2$, a Gaussian function centered at $x=y=0$ with waist radius of $12\mu$m, and therefore with $L_{diff}=1.6$mm.

\h The initial fields of $\Psi_1$ and $\Psi_2$ are used in evaluating the solutions of Eq.(\ref{NLS}). Figure (2a) shows the intensity of the resulting FW beam (a slice at the $y=0$ plane) in the nonlinear medium, where it can be seen that the FW can resist the effects of diffraction and nonlinearity, maintaining the desired light structured intensity pattern. Figure (2b) illustrates the behavior of the Gaussian beam (within the hollow FW), clearly demonstrating its guided propagation in the regions where the FW is activated (on). As expected, the Gaussian beam undergoes diffraction in the regions between the cylinders where the FW is deactivated (off), as well as upon exiting the final light cylinder. The subfigure inset in Fig. 2(b) depicts the on-axis intensity profile of the Gaussian beam throughout its propagation.

\

{\bf A taper shaped guiding structure made of light}

\h Here, we will utilize an optical beam structured as a taper to guide and compress the Gaussian beam traveling through it. To create this light-based guide, we will employ the linear FW from the second example in Section 2, but here we choose an increasing amplitude along the propagation determined by $f(z) = 2\sqrt{3}A_g \exp\left(\frac{2.6z}{L}\right)$ from $z=0$ to $z=L/2=10$cm, where $A_g$ is the initial amplitude of the Gaussian beam. As before, we set $z=0$ in the resulting linear FW and use this as the initial condition for $\Psi_1$ (the light-based guide) in the nonlinear medium. The initial field $\Psi_2$, which will be injected into the structured beam $\Psi_1$, is represented by a Gaussian function with a waist radius of $55\mu$m (thus yielding $L_{diff}=3.38$cm), where its amplitude is the previously defined, $A_g$.

\h The initial fields $\Psi_1(x,y,z=0)$ and $\Psi_2(x,y,z=0)$ are employed in the numerical solution of Eqs.(\ref{NLS}). In Fig.(2c), the resulting intensity pattern of the FW structured beam in the nonlinear medium is displayed over the $y=0$ plane, exhibiting the desired taper shape. Figure (2d) shows the Gaussian beam that has been inserted into the structured beam (FW), where it is guided and compressed by the light-base guide. The subfigure inset in Fig.(2d) presents the on-axis intensity profile of the guided Gaussian beam. It is evident that once the Gaussian beam exits the FW, it is no longer guided and suffers the effects of diffraction.

\

{\bf A guiding and splitting structure made of light}

\h Using our approach, next we will construct a hollow Y-splitter made of light that guides and splits a Gaussian beam inserted into it. The construction of the desired light-based waveguide in the nonlinear medium starts by its construction in the linear one (with refractive index $n_l=1.3$). In this process, three hollow FWs of fourth-order are utilized (all with $L=10$cm); the first FW, with radius of $26.5\mu$m (which corresponds to a FW's Q parameter of $0.99992k_1$) is initiated at $z=0$ and propagates along the $z$-direction with a constant peak amplitude of $3.6A_g$ until reaching approximately $2.5$cm. At this point, this structured beam connects with two new fourth-order FWs with radii of $24.9\mu$m (with $g(z)=0$ and FW's Q parameters of $0.99991k_1$), constant amplitudes of $3.9A_g$ and propagating along directions shifted by $\pi/800$ radians above and below the $z$-axis (the three propagation axes lie on the plane $y=0$), covering a distance of approximately $2.5$cm before being, by our choice, "turned off" (i.e., before being spatially dissolved). The resulting analytical solution, made from the superposition of these three FWs in the linear medium, is evaluated at $z=0$ as the initial $\Psi_1$ beam in the linear medium.

\h The initial field of $\Psi_2$ (the Gaussian beam that will be injected into the hollow structured beam) is given by a Gaussian function in the $z=0$ plane, centered at $x=y=0$, of amplitude $A_g$ mentioned previously, and with waist radius of $20\mu$m, and consequently $L_{diff}=4.47$mm. The initial fields $\Psi_1(x,y,z=0)$ and $\Psi_2(x,y,z=0)$ are used in the numerical solution of Eqs.(\ref{NLS}).

\h Figure (2e) depicts the intensity distribution pattern of the structured beam $\Psi_1$ on the plane $y=0$, highlighting its desired spatial shape in the nonlinear medium. Within the same plane, Fig. (2f) shows the intensity of the Gaussian beam that is initially guided by the first FW, and subsequently split into two separate beams, which are guided by the second and third FWs. Eventually, at the termination of the guiding region, the Gaussian beams undergo diffraction.

\

{\bf A guiding and combining structure made of light}

\h The guiding structure in this example is essentially a reversed version of the previous example, functioning as a combiner. To construct it in the linear medium with $n_l=1.3$, we utilize two hollow FWs of the fourth order with radii of $37.5\mu$m, corresponding to $g(z)=0$ and a FW's Q parameter of $0.99996k_1$. Their propagation axes lie on the $y=0$ plane and originate from points $(-100\mu\rm{m},0,0)$ and $(100\mu\rm{m},0,0)$, with inclinations of $+\pi/800$ radians and $-\pi/800$ radians with respect to the $z$-axis. The two FWs propagate with constant amplitudes of $3A_g$ over a distance of approximately 2 cm, after which they connect with a third FW of the fourth order, which also has a radius of $37.5\mu$m (resulting in $g(z)=0$ and the same $Q$ parameter as the previous two FWs) and propagates with a constant amplitude of $6A_g$ parallel to the $z$-axis, covering a distance of approximately 2.55 cm. For these three FWs, the parameter $L$ was set to 30 cm for the first two and 20 cm for the third. As usual, the resulting optical field, obtained by superposing these three FWs in the linear medium, is evaluated at $z=0$ and considered as the initial field for $\Psi_1$ in the nonlinear medium.

\h Representing the initial field of $\Psi_2$, i.e., the two Gaussian beams that will be injected into the two inclined FWs, we consider two Gaussian functions with a waist radius of $20\mu$m, corresponding to $L_{diff}=4.47$mm. These Gaussians are centered at the points $(-100\mu\rm{m},0,0)$ and $(100\mu\rm{m},0,0)$, both with amplitude $A_g$. The initial fields $\Psi_1(x,y,z=0)$ and $\Psi_2(x,y,z=0)$ are employed in the numerical solution of the Eqs.(\ref{NLS}).

\h Figure (2g) displays, on the plane $y=0$, the intensity of the resulting structured field $\Psi_1$ in the nonlinear medium, demonstrating the successful construction of the desired spatial structure. Additionally, on the same plane, Fig. (2h) exhibits the intensity profiles of the two Gaussian beams guided by the initial two FWs and subsequently combined into a single beam within the third FW. The combined beam continues to be guided by the third FW along its entire path until it eventually undergoes diffraction upon exiting the light-based guide.

\begin{figure}[ht!]
\centering\includegraphics[width=16cm]{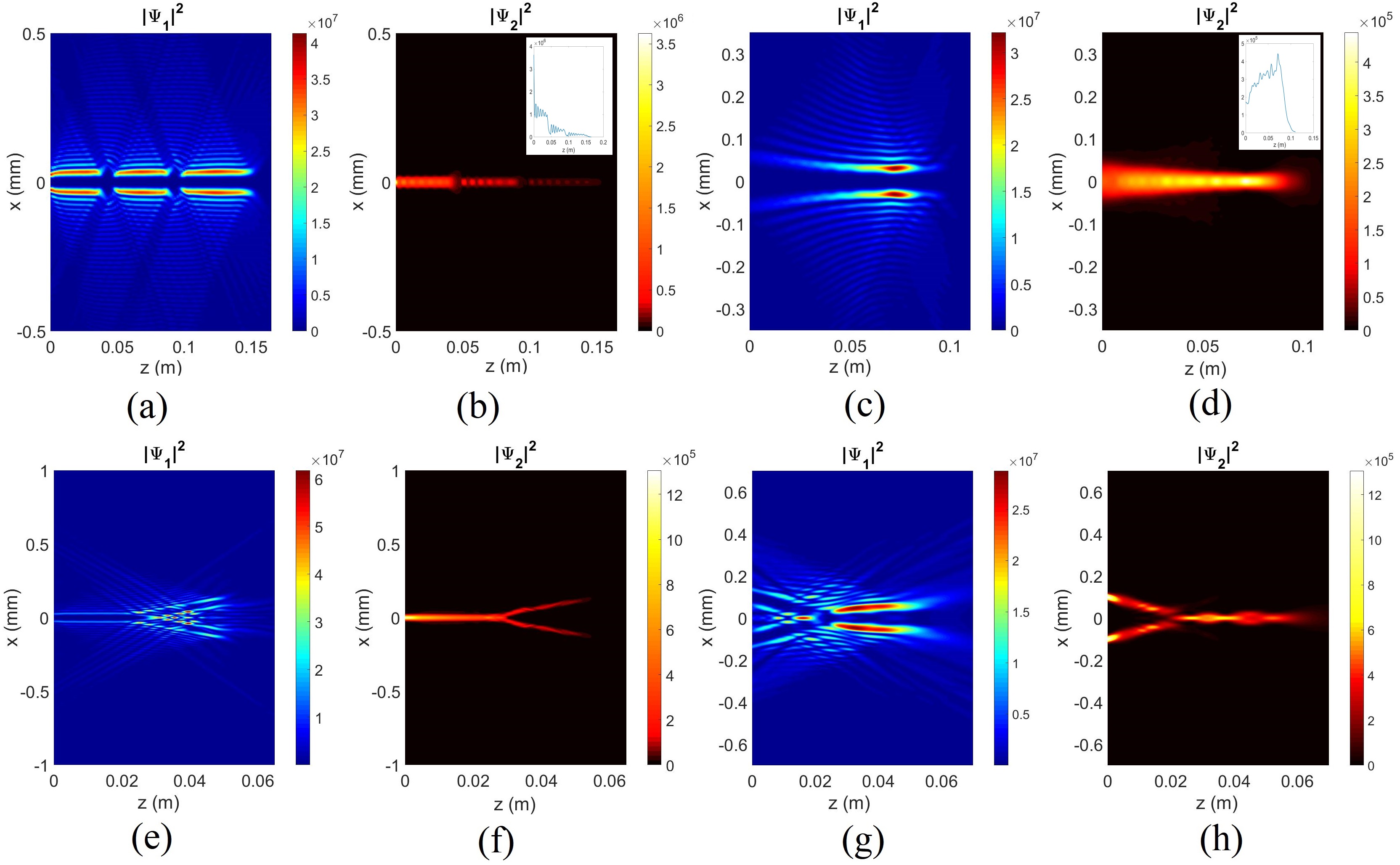}
\caption{The figures display, in pairs, the guides created from structured light and their respective guided Gaussian beams in a bulk self-defocusing Kerr medium with $n_2 = -10^{-9} \rm{m}^2/\rm{W}$, specifically: a) the hollow guide-based light in a triple on-off configuration, and b) the respective guided Gaussian beam, with the subfigure inset depicting its on-axis intensity profile throughout propagation; c) a taper-shaped-hollow guide made of structured light, and d) the respective guided and compressed Gaussian beam, with the subfigure inset depicting its on-axis intensity profile throughout propagation; e) a hollow Y-splitter made of structured light, and f) the respective Gaussian beam being guided and subsequently split into two guided beams; g) a hollow combiner made of structured light, and h) the respective pair of Gaussian beams being guided and subsequently combined into a guided single beam.}
\end{figure}

\subsection{An example for $n_2>0$.}

{\bf An angled guiding structure made of light}

\h For the last example, we will construct an angled guiding structure (a bend waveguide) using FWs in a Kerr medium with positive $n_2$. Similar to the previous cases, we start by designing the structured beam in the linear medium with $n_l=1.3$. We employ two zero-order FWs whose propagation axes lie on the $y=0$ plane, each with a spot radius of $17\mu$m (thus, for both FWs, we have $g(z)=0$ and $Q=0.99996k_1$). The first FW starts from the coordinate system's origin and propagates a distance of $4.5$cm along the $z$-direction with an amplitude described by the morphological function $F(z)= \sqrt{7}A_g(0.3 + \exp(6.66z))$. After this distance, the first FW waveguide connects with the second FW, which propagates another $4.5$cm in a direction deviated by $\pi/800$ radians from the $z$-axis with an amplitude given by the morphological function $f(z')= 1.5\sqrt{7}A_g(0.3 + \exp(6.66z'))$, where $z'$ represents the new propagation axis. For both beams, we set the parameter $L$ equal to $30$cm. As before, the resulting optical field, obtained by superposing the two FWs in the linear medium, is evaluated at $z=0$ and considered as the initial field for $\Psi_1$ in the nonlinear medium. On the other hand, for the initial field of $\Psi_2$ (the Gaussian beam injected into the first FW), we have a Gaussian function with a waist radius of $20\mu$m, corresponding to $L_{diff}=4.47$mm, centered at the origin and possessing the amplitude $A_g$. The initial fields $\Psi_1(x,y,z=0)$ and $\Psi_2(x,y,z=0)$ are employed in the numerical solution of the Eqs.(\ref{NLS}).

\h Figure (3a) depicts the intensity distribution of the structured beam $\Psi_1$ on the $y=0$ plane, where we can see it possesses the desired spatial shape. Within the same plane, Fig.(3b) shows the intensity of the Gaussian beam that is initially guided by the first FW, and subsequently shifted and guided by the second FW. Eventually, at the termination of the guiding region, the beam experiences diffraction.

\begin{figure}[ht!]
\centering\includegraphics[width=13cm]{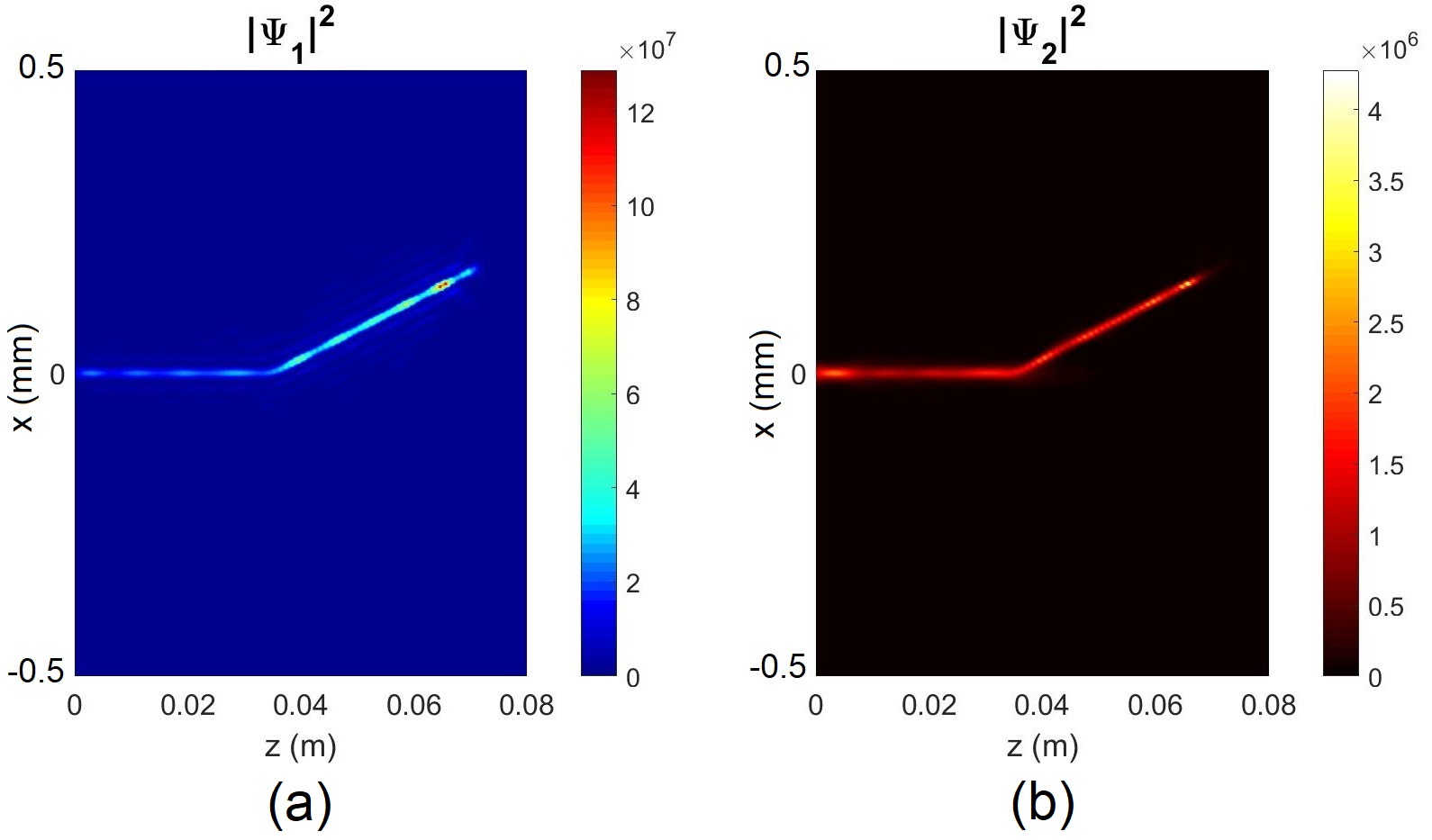}
\caption{Example considering a bulk self-focusing Kerr medium with $n_2 = 10^{-9} \rm{m}^2/\rm{W}$: a) an angled guiding made of structured light, and b) the guided Gaussian beam.}
\end{figure}

\section{Conclusions}

We have constructed structured FW type optical beams with a variety of geometries in Kerr nonlinear media, both with negative and positive $n_2$ coefficients. Such structured lights were utilized as guiding structures, tapers, beam splitters, and beam combiners for conventional Gaussian beams. Our results reinforce that employing structured light in nonlinear media can be of great interest in light controlling light and for all-optical photonics applications.

\section*{Acknowledgments}
Thanks are due to partial support from S\~ao Paulo Research
Foundation (FAPESP) (under grant 2021/15027-8) and from CNPq
(under grant 305475/2022-3). This work  was also supported by Natural Sciences and Engineering Research Council of Canada (NSERC).


\begin{thebibliography}{1}

\bibitem{livro1} Non-Diffracting Waves, edited by H.E.Hern\'andez-Figueroa, E.Recami, and M.Zamboni-Rached
(J.Wiley; Berlin, 2014).

\bibitem{livro2} Structured Light and Its Applications: An Introduction to Phase-Structured Beams and Nanoscale Optical Forces,
edited by David L. Andrews (Academic Press, Cambridge, Mass., 2008).

\bibitem{fw1} M.Zamboni-Rached, ``Stationary optical wave fields with arbitrary longitudinal shape by superposing equal
frequency Bessel beams: Frozen Waves,'' Opt. Express  $\mathbf{12}$(17), 4001--4006 (2004).

\bibitem{fw1b} Michel Zamboni-Rached, ``Diffraction-Attenuation resistant beams in absorbing media,'' Opt. Express 14, 1804-1809 (2006).

\bibitem{fw1c} Tarcio A. Vieira, Marcos R. R. Gesualdi, and Michel Zamboni-Rached, ``Frozen waves: experimental generation,'' Opt. Lett. 37, 2034-2036 (2012).

\bibitem{fw1d} Ahmed H. Dorrah, Michel Zamboni-Rached, and Mo Mojahedi, ``Generating attenuation-resistant frozen waves in absorbing fluid,'' Opt. Lett. 41, 3702-3705 (2016).

\bibitem{feixe_curvo} Michel Zamboni-Rached, ``Simple and analytical method for controlling the trajectory and branching of optical beams,'' J. Opt. Soc. Am. B 38, 448-455 (2021)

\bibitem{yousuf1} Yousuf Aborahama, Ahmed H. Dorrah, and Mo Mojahedi, ``Designing the phase and amplitude of scalar optical fields in three dimensions,'' Opt. Express 28, 24721-24730 (2020).

\bibitem{yousuf2} Yousuf Aborahama, Rajat K. Sinha, and Mo Mojahedi, ``Customized Vectorial Optical Fields in Homogeneous and Inhomogeneous Media,'' Phys. Rev. Applied 18, L031002 (2022).

\bibitem{fw2} M. Corato-Zanarella, A. H. Dorrah, M. Zamboni-Rached, and M. Mojahedi, ``Arbitrary control of polarization and intensity profiles of diffraction-attenuation-resistant beams along the propagation direction,'' Phys. Rev. Appl. 9(2), 024013 (2018).

\bibitem{forbes2} A. H. Dorrah, C. Rosales-Guzmán, A. Forbes and M. Mojahedi, ``Evolution of orbital angular momentum in three-dimensional structured light,'' Phys. Rev. A 98, 043846 (2018).

\bibitem{ahmed} A. H. Dorrah, M. Zamboni-Rached, M. Mojahedi, Experimental demonstration of tunable refractometer based on orbital angular momentum of longitudinally structured light. Light Sci. Appl. 7, 40 (2018).

\bibitem{pinca} Rafael A. B. Suarez, Leonardo A. Ambrosio, Antonio A. R. Neves, Michel Zamboni-Rached, and Marcos R. R. Gesualdi, ``Experimental optical trapping with frozen waves,'' Opt. Lett. 45, 2514-2517 (2020).

\bibitem{fw3} Michel Zamboni-Rached, ``Carving beams of light,'' Opt. Lett. 46, 1205-1208 (2021).

\bibitem{livro3} Jerome V. Moloney, Alan C. Newell, Nonlinear Optics (CRC Press, NW, 2018).

\bibitem{forbes} Buono WT, Forbes A. Nonlinear optics with structured light. Opto-Electron Adv 5, 210174 (2022).

\bibitem{Bessel_NL} A. Balbuena Ortega, F. E. Torres-Gonz\'alez, V. L\'opez Gayou, R. Delgado Macuil, J. E. H. Cardoso Sakamoto, A. V. Arzola, G. Assanto, K. Volke-Sepulveda, ``Guiding light with singular beams in nanoplasmonic colloids,'' Appl. Phys. Lett. 8 February 2021; 118 (6): 061102.

\bibitem{valor_n2} A. Balbuena Ortega, E. C. Brambila, V. L\'opez Gayou, R. Delgado Macuil, A. Ordu\~na Diaz, A. Zamilpa Alvarez, A. V. Arzola and K. Volke-Sep\'ulveda, ``Light control through a nonlinear lensing effect in a colloid of biosynthesized gold nanoparticles,'' Journal of Modern Optics, 66:5, 502-511 (2019).

 \end{thebibliography}
\end{document}